\documentclass[%
 aip,
 jmp,%
 amsmath,amssymb,
 reprint,%
]{revtex4-2}

\usepackage{graphicx}
\usepackage{dcolumn}
\usepackage{bm}

\usepackage{color}
\newcommand{\change}[1]{{{#1}}}

\begin{document}

\preprint{AIP/123-QED}

\title{Confinement enhanced viscosity vs shear  thinning in lubricated ice friction}

\author{ {\L}ukasz Baran }
\affiliation{ 
Department of Theoretical Chemistry, Institute of Chemical Sciences, Faculty of
Chemistry, Maria Curie-Sklodowska University in Lublin, Lublin, Poland
}%
\author{Luis G. MacDowell}%
 \email{lgmac@quim.ucm.es}
 \affiliation{Departamento de Qu\'{i}mica F\'{i}sica, Facultad de Ciencias Q\'{i}micas,
 Universidad Complutense de Madrid, 28040, Spain.}%

\date{\today}

\begin{abstract}
The ice surface is known for presenting a very small kinetic friction coefficient, but the origin of this property remains highly controversial to date. In this work, we revisit recent computer simulations of ice sliding on atomically smooth substrates, using newly calculated bulk viscosities for the TIP4P/Ice water model. The results show that spontaneously formed premelting films in static conditions exhibit an effective viscosity which is about twice the bulk viscosity. However, upon approaching sliding speeds in the order of m/s, the shear rate becomes very large, and the viscosities decrease by several orders of magnitude. This shows that premelting films can act as an efficient lubrication layer despite their small thickness, and illustrates an interesting interplay between confinement enhanced viscosities, and shear thinning. Our results suggest that the strongly thinned viscosities that operate under the high speed skating regime could largely reduce the amount of frictional heating.
\end{abstract}

\pacs{Valid PACS appear here}
\maketitle



The origin of ice slipperiness is a long standing problem that is attracting renewed interest.\cite{weber18,canale19,liefferink21,lever21b,du23b,persson23,miyashita23} 
The classical interpretation of this phenomena emphasizes the role of melt-water, originated from frictional heating, as the reason for the low kinetic friction coefficient of ice. This idea, which has an old history,\cite{bowden53} was developed in quantitative form already a while ago,\cite{oksanen82,colbeck88} and has been refined
over the years to  account for surface roughness, abrasion and
wear.\cite{lozowski13,lever23,du23} However, the significance of frictional melting has been recently
challenged,\cite{tusima11,weber18,canale19,liefferink21,lever21b}  while other mechanisms consistent with lubrication, such as pressure melting, or spontaneous ice premelting, 
have received far less attention, despite evidence of equilibrium interfacial premelting in a variety of substrates.\cite{beaglehole94,strausky98,liljeblad17,ribeiro21} 

Recently, we studied the role of spontaneous interfacial premelting in ice friction by means of computer simulations.\cite{baran22} For the TIP4P/Ice model,\cite{abascal05} our study shows that at temperatures above -20 $^\circ$C, equilibrium premelting films in the order of one nanometer are formed spontaneously and can act as an efficient lubrication layer.
In fact, based on our results, we found that, at sliding speeds in the scale of m/s, the rheology of the premelting film could be described with a model of Couette flow with slip, together with a shear rate  dependent viscosity 
to account for shear thinning. Accordingly, the shear stress could be quantified as:
 \begin{equation}\label{eq:model}
 \tau = \frac{\eta(\dot{\gamma}) U}{d_C + b}
 \end{equation}
 where, $\eta(\dot{\gamma})$ is the viscosity, $\dot{\gamma}$ is the shear rate,
$U$ is the slider speed, {$d_C$} is the Couette flow thickness that accounts for a small negative slip at the ice/film surface,\cite{louden17} and $b$ is a  substrate dependent slip length.

By coincidence, our results, performed mainly at 262~K, and a sliding speed of $U=5$~m/s, provided hydrodynamic viscosities $\eta(\dot{\gamma})$, somewhat larger, but   similar, to the viscosity of bulk undercooled water, which is unexpected as confinement can increase significantly the effective viscosity.\cite{robbins00}
However, {for a slider shearing a lubricating premelting film,} the
shear rate is given by $\dot{\gamma}\approx U/d_C$. {Since the equilibrium films are barely 1~nm thick, it suffices a sliding speed in the scale of m$/$s, to achieve rates of the order
$\dot{\gamma}\approx 10^9$~s$^{-1}$, well beyond the upper limit of current state of the art rheometry.\cite{pipe08}} {This rate is sufficient
to trigger shear thinning even for a molecular fluid such as water,\cite{ribeiro20} which
is the paradigmatic example of a 'simple' Newtonian liquid.\cite{pipe08} Therefore,
the occurrence of shear thinning  in our simulations}  could obscure the expected confinement induced enhancement of the film's viscosity.


Indeed, by studying interfacial sliding in a lower range of temperatures  between 230 and 250~K, we found clear hints of shear thinning. \change{This could be confirmed by  explicitly shearing the interfacially premelted films, and calculating  hydrodynamic viscosities from the ratio $\eta(\dot{\gamma})=\tau/\dot{\gamma}$, with $\tau$ measured as the shear force the substrate exerts on the water molecules, and $\dot{\gamma}$, from the slope of the flow profile.\cite{baran22}} The results were one to two orders of magnitude smaller than the {Green-Kubo} viscosities of undercooled water at the corresponding temperature,\cite{baran22} and lead us to suggest interfacially premelted films in ice displayed essentially bulk-like behavior, with a shear rate dependent viscosity consistent with the Eyring model of shear thinning:\cite{spikes14}


\begin{equation}\label{eq:eyring}
 \eta(\dot{\gamma}) = \frac{\tau_0}{\dot{\gamma}} \sinh^{-1}\left(\frac{\eta_0\dot{\gamma}}{\tau_0} \right)
\end{equation}
Here, $\eta_0$ is the bulk viscosity at zero shear rate, and $\tau_0\propto T$ is a threshold shear stress above which shear thinning becomes significant. 

\begin{figure}
	\includegraphics[width = 0.5\textwidth]{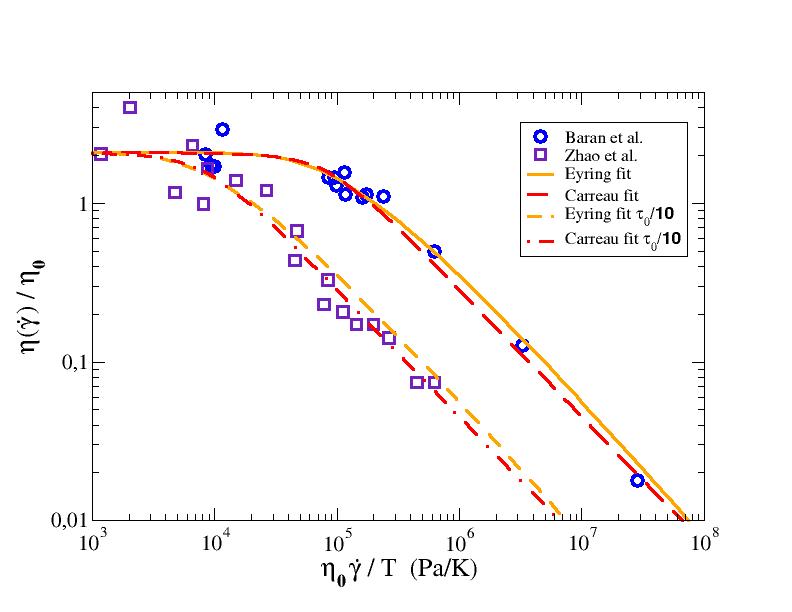}
	\caption{\label{fig:thinning} Viscosity of premelted liquid films at the
ice/substrate interface. Plot shows computer simulation results from Baran et
al.\cite{baran22} (circles) and Zhao et al.\cite{zhao22} (squares). Lines
\change{across}
the Baran et al. data \change{are fits} to the Eyring and Carreau models of shear
thinning. Notice the fits plateau to a value corresponding to twice the bulk
viscosity. \change{Lines across the data of Ref.\cite{zhao22} are the same Eyring
and Carreau fits, albeit with $\tau_0$ decreased by a factor of 10.}
}
\end{figure}

This conclusion was reached in Ref.\citep{baran22}, based on a subset of the data, corresponding to simulations performed for ice sliding at $U=5$~m/s with pressure $p=0.1$~MPa over a hydrophilic slider  with contact angle $\theta=50^\circ$ (c.f. Fig.S10 of Ref.\cite{baran22}). Upon revisiting our own published data for a wider range of conditions(c.f. Tables S1 to S3 of Ref.\cite{baran22}), including results at $p=60$~MPa, and sliding speed $U=0.5$~m/s, for both hydrophilic and hydrophobic sliders, we find that our results actually show an interesting interplay between enhanced viscosity due to confinement,\cite{robbins00} and depleted viscosities due to shear thinning.\cite{ribeiro20} 

This can be illustrated by plotting the ratio $\eta(\dot{\gamma})/\eta_0$, as a function of $\eta_0\dot{\gamma}/T$,  which, according to Eq.\ref{eq:eyring}, should exhibit a smooth decay in a log-log scale; c.f. Figure~(\ref{fig:thinning}). A plateau value at small shear rate, followed by an exponential fall at increasing
shear rate is the {hallmark} of shear thinning anticipated in Ref.\cite{baran22}. However, it is noticed that the results at low shear rate do not converge to unity, but instead are scattered about  a plateau value  roughly twice as large. This suggests that, in the limit of zero shear rate, the confinement of the premelting layers results in an overall increase of the effective viscosity by a factor of two. 


This effect can be crudely estimated by replacing $\eta_0$ with $f \eta_0$ in the 
Eyring equation, Eq.(\ref{eq:eyring}), using $f>1$ as an enhancement parameter to account for the
increase of the zero shear rate viscosity of the premelting film with respect to
the bulk value, $\eta_0$.  The fit to this model, illustrated as an orange line
in Fig.~(\ref{fig:thinning}), shows good agreement with our data for a
threshold $\tau_0/T = 9.13\cdot 10^4$~Pa/K, and $f=2.10$.

\change{Remarkably, the modified Eyring equation displayed in Fig.1 implies that the ratio $\eta(\dot{\gamma})/\eta_0$ for both hydrophilic and hydrophobic substrates at different sliding speeds, pressures and temperatures (four variables) all collapse roughly into a (single variable) master function of the Peclet number, i.e. the ratio of viscous to diffusive flow.}
Indeed, using the Stokes-Einstein relation, the Peclet number for flow in the scale of a molecular diameter, {$\change{a}$}, is given by $Pe=\pi \change{a}^3\eta_0 \dot{\gamma}/2k_BT$, which is equal to the single variable $\frac{\eta_0\dot{\gamma}}{T}$ up to a constant factor of $\tau_0 = k_BT/v^{\ddagger}$, where $v^{\ddagger}$ is an activation volume.
{For the above fit, this corresponds to a meaningful activation volume $v^{\ddagger}=6.7\cdot 10^{-28}$~m$^3$ that falls in the expected molecular scale.} These results confirm 
the significance of shear thinning reported in Ref.\cite{baran22}, as well as the 
confinement induced enhancement of the viscosity in the limit of vanishing shear rate.

Alternatively to the Eyring model, the shear thinning can be also modeled with a modified Carreau equation:\cite{spikes14}
\begin{equation}\label{eq:carreau}
\eta(\dot{\gamma}) =   {  {f \eta_0} \bigg / {
		\left ( 1 + \left( \frac{f \eta_0\dot{\gamma}}{\tau_0}
\right)^{\change{\alpha}}  \right )^{\frac{1-n}{\change{\alpha}}} }
}
\end{equation}
with the enhancement factor, $f$, $\tau_0$ and the exponents $\change{\alpha}$ and $n$ used as fitting parameters. 
A best fit, using $\change{\alpha}=2$, and  $(1-n)/\change{\alpha}=0.4$, provides $\tau_0/T=1.67\cdot 10^5$~Pa/K and $f=2.08$, 
in good agreement with the threshold values for shear thinning of bulk
water,\cite{ribeiro20}
and consistent with an increase 
of the premelting film viscosity $\eta(0)$ by a factor of two with respect to the bulk value $\eta_0$. 
Particularly, our fit provide $\tau_0$ of $3.8\cdot 10^7$, $4.2\cdot 10^7$ and $4.4\cdot 10^7$ Pa,
for $T=230, 250$ and 266~K, respectively. \change{This is not far from} the data of Ref.\cite{ribeiro20}
for bulk water, which yield $3.3\cdot 10^7$, $6.0\cdot 10^7$ and $7.4\cdot 10^{7}$~Pa for
the corresponding temperatures, \change{and illustrates the similarity of the
premelting film's rheology with bulk water.}

In a subsequent study, Zhao et al.\cite{zhao22} performed computer simulations
for the same model{, shearing  the ice surface with sliding speeds ranging
from 0.1 to 30~m/s. At a temperature of 240~K,  their hydrodynamic viscosities
decreased from about 29~mPa$\cdot$s at low share rate to about 0.9~mPa$\cdot$s at high shear rate, which appears indicative of shear thinning. 
Unfortunately, the authors compared their data with a
bulk water viscosity $\eta_0=0.8$~mPa$\cdot$s  measured at a density of 1~g/cm$^3$ and $T=300$~K.\footnote{The thermodynamic state of the reference viscosity used in the work of \citep{zhao22}  was not reported in the paper, but communicated to us privately.} 
With this gauge, the hydrodynamic viscosity of the confined films at equilibrium would appear to be more than 30 times larger than $\eta_0$, and would decrease down to the bulk viscosity $\eta_0$ only at large shear rate. This appears to imply the confinement effect prevents the films to exhibit viscosities smaller than $\eta_0$, at odds with expectations for plain shear thinning in bulk.  To clarify this issue, we have calculated bulk viscosities of the TIP4P/Ice model at the conditions relevant to the study of Zhao et al. by the Green-Kubo method, following the methodology reported in Ref.\citep{baran23}. At the pressure of  $p=133$~MPa, our results provide bulk viscosities of $12.7\pm0.4$, $5.5\pm0.1$ and $3.73\pm0.06$ mPa$\cdot$s, for the temperatures of 240, 255 and 265~K, respectively.
}

The calculated bulk viscosities now allow us to assess the role of confinement and shear thinning in the  data of Ref.\cite{zhao22}. To show this, we display the ratio of $\eta(\dot{\gamma})/\eta_0$, using $\eta(\dot{\gamma})$ from Fig.8 of Zhao et al. as a function of $\eta_0\dot{\gamma}/T$, with the shear rate estimated roughly as $\dot{\gamma}=U/\delta$, and the film thickness $\delta$, as displayed in Fig.5 of their paper. Similar to our own work, the results appear to be roughly consistent with a moderate enhancement of the viscosity by a factor of about 2, {and  allow us to confirm the decrease of the hydrodynamic viscosity well below $\eta_0$, as expected for a case of shear thinning.\cite{baran22}} However, the onset of shear thinning appears to start earlier. {In principle,} this discrepancy could be related to differences in the model substrates used (a BCC lattice in Ref.\cite{zhao22} and an FCC lattice in Ref.\cite{baran22}). {However, note that Zhao et al. did not actually simulate ice Ih, but a metastable hydrogen ordered analog with orthorombic symetry (P222). Their simulations were carried out at a pressure of 133 MPa, that for the highest temperature of their study, $T=265~K$, is about twice the bulk melting point of the TIP4P/Ice.\cite{abascal05,baran22} Furthermore, their system was equilibrated over barely
of 0.6~ns,\footnote{The timestep of the simulations was not reported in
Ref.\cite{zhao22}. Real time estimates here are obtained using $dt=1$~fs, as
communicated to us privately by the authors.} while the relaxation time of
interfacial ice premelting is in the order of decades of
nanoseconds.\cite{baran22} Despite these shortcomings, the results of Zhao et al. 
\change{may be described with exactly the same Eyring and Carreau fits} as our work, albeit with a threshold shear stress $\tau_0$ a factor of 10 times smaller.}

In summary, current computer simulations of the TIP4P/Ice model support the
significance of surface premelting in lubricated ice friction. The results
suggest that confinement increases the zero shear rate viscosity of the
premelting films by a small factor of less than one order of magnitude. However,
the onset of shear thinning that takes place for shear rates in the scale of
$10^7$-$10^9$ Hz, can decrease the hydrodynamic viscosity by orders of
magnitude. This behavior, which has not been recognized until
recently,\cite{baran22} could play a very important role in the account of ice slipperiness and limit the extent of frictional melting. The presence of shear thinning also helps explaining why the dependence of the friction coefficient with sliding speed occurs in the $\ln U$ scale,\cite{liefferink21,miyashita23} and flattens at large sliding speed.\cite{zhao22}

\section*{AUTHOR DECLARATIONS}

\subsection*{Conflict of interest}

The authors have no conflicts to disclose.
\\

\subsection*{Author Contributions}

L. Baran performed calculations; L.G. MacDowell designed research and drafted note. All authors participated in discussions.

\section*{DATA AVAILABILITY}

The Data that support the findings of this study are available within the article or in the listed references \cite{baran22,zhao22}.

\begin{acknowledgments}
	We acknowledge funding from the Spanish Agencia Estatal de Investigaci\'on
	under research grant PID2020-115722GB-C21/AEI/10.13039/501100011033.
We benefited from generous allocation of computer time at the Academic Supercomputer Centre (CI TASK) in Gdansk.
\end{acknowledgments}

\bibliography{/home/luis/Ciencia/tex/Patrones/referenc}

\end{document}